\documentstyle[preprint,aps]{revtex}
\begin{document}
\draft
\title{The Anderson Model out of equilibrium: Time dependent
perturbations}
\author{Matthias H. Hettler$^1$ and Herbert Schoeller$^{2,}$$^3$}
\address{$^1$Department of Physics and National High Magnetic Field Laboratory,
University of Florida, Gainesville, FL 32611,
e-mail: hettler@neptune.phys.ufl.edu}
\address{$^{2,}$ Institut f\"ur theoretische Festk\"orperphysik,
Universit\" at Karlsruhe, 76128 Karlsruhe, Germany}
\address{$^3$ Department of Physics, Simon Fraser University,
Burnaby, B.C. V5A 1S6, Canada, e-mail: hschoell@sfu.ca}
\date{Received \today}
\maketitle
\begin{abstract}
The influence of high-frequency fields on quantum transport through a
quantum dot is studied in the low-temperature regime. We generalize
the non crossing approximation for the infinite-U Anderson model to
the time-dependent case. The dc spectral density shows asymmetric
Kondo side peaks due to photon-assisted resonant tunneling. As a
consequence we predict an electron-photon pump at zero bias which is
purely based on the Kondo effect. In contrast to the resonant level
model and the time-independent case we observe asymmetric peak
amplitudes in the Coulomb oscillations and the differential
conductance versus bias voltage shows resonant side peaks with a width
much smaller than the tunneling rate. All the effects might be used to
clarify the question whether quantum dots indeed show the Kondo effect.

\end{abstract}

The influence of time-dependent external fields on quantum transport
through mesoscopic devices has attracted a lot of interest in the last
few years. In the noninteracting case ac response of resonant
tunneling devices has been studied both experimentally and
theoretically by many authors
\cite{win-jau-mei,joh,che-tin,liu,run-ehr}. Recently
Coulomb interactions have been included by B\"uttiker et.al.
\cite{bue} by a self-consistent potential method to account correctly
for the displacement currents. In addition, for ultrasmall metallic tunnel
junctions and quantum dots in the nanoscale regime, Coulomb blockade
phenomena are important. They have been studied in the presence of
time-dependent fields either for low \cite{pot-laf,kou-joh,odi} or
high frequencies \cite{dev-lik,bru-sch,kou-jau}. Interesting
effects like single-electron pumps \cite{pot-laf}, turnstiles
\cite{kou-joh,odi} and
electron-photon pumps \cite{bru-sch} have been predicted. Furthermore,
the possibility of photon-assisted tunneling in connection with
single-electron effects gives rise to a multitude of new resonant
features in the Coulomb oscillations and the Coulomb staircase
\cite{bru-sch,kou-jau}. Some of these effects have been
observed in a recent experiment by Kouwenhoven et.al. \cite{kou-jau}
by applying microwave irradiation of 19GHz to a split-gate quantum dot
device. However, all these papers deal with the high-temperature
regime where the classical picture of sequential tunneling applies.

In this letter we will investigate for the first time the influence of
high-frequency fields on single-electron effects for very low
temperatures where coherent resonant tunneling processes become
important. To display the physical effects occurring in such a case
most clearly we concentrate here on a quantum dot with two degenerate
energy levels subject to an explicit time-dependent external field.
For large Coulomb-interaction we can represent this system by the
infinite-U Anderson model with spin degeneracy $N=2$. The time
dependent modulation can either be applied to the electrons in the
leads or to the two levels on the quantum dot. In the time-independent
case this model has already attracted much interest since the
occurrence of a Kondo resonance at low temperatures increases the
conductance of the system considerably
\cite{ng-lee,gla-rai}. Furthermore the differential conductance versus
bias voltage shows a sharp maximum for zero bias due to a splitting
of the Kondo resonance \cite{her-dav-wil,mei-win-lee}. The
research in this direction is further motivated by a recent experiment
of Ralph \& Buhrman \cite{ral-bur}. At very low temperatures they
found singularities in the differential resistance of metal point
contacts containing defects which can be modeled by a two level
system or a two channel Anderson model \cite{het-kro-her}. The
experimental data seem to be in agreement with theoretical
investigations, either using equilibrium conformal field theory
\cite{ral-lud} or nonequilibrium Green functions
\cite{het-kro-her}.

In the presence of time-dependent perturbations we will combine in
this work the picture of photon-assisted tunneling according to Tien \&
Gordon \cite{tie-gor} with resonant tunneling phenomena and the Kondo
physics. The condition for the Kondo resonance to occur is a very low
temperature to enhance the sharpness of the Fermi surface which gives
rise to a sharp peak of the spectral density near the Fermi energy. If
we apply the external fields only to the leads,
the electrons there will absorb or emit multiples of the photon
energy $\hbar\Omega$, where $\Omega$ is the applied external frequency.
As a consequence, there will be a whole set of sharp Fermi surfaces in the
reservoirs. Thus in the low temperature regime the Kondo peak will
split up into a whole set of Kondo peaks, separated by the external
frequency and weighted approximately by $J_n^2({\Delta\over\Omega})$,
where $J_n$ denotes the Bessel function of order n and $\Delta$ is the
power of the applied external field. This leads to a drastic change of
the spectral density of the dot due to ac-voltages in the leads which
can not be observed in the case of a single resonant level. In the
latter case the only effect is photon-assisted tunneling due to
excitations of the electrons in the leads but the spectral density in
the dot remains completely unchanged. We will calculate and analyze these
new features by generalizing the Non Crossing Approximation (NCA) for
nonequilibrium systems to the present case. Our detailed numerical
analysis shows that the distribution of the Kondo peak amplitudes is
significantly asymmetric and depends on the value of the gate voltage.
For an asymmetric coupling of the external fields to the two leads
this gives rise to a pump effect at zero bias which is solely due to
the Kondo physics. Furthermore we predict asymmetric
side peaks in the Coulomb oscillations and find resonant peaks in the
differential conductance versus bias voltage which are due to overlaps
of Kondo peaks arising either from the finite voltage or the external
field. All the effects can not be observed in the time-independent
situation or for the $N=1$ case with only one resonant level. Thus our
results may help to find further experimental evidences for the
occurence of Kondo phenomena in mesoscopic devices.

The system under consideration is an Anderson  hamiltonian
of a particle on a dot level with infinite strong
Coulomb repulsion coupled to two large electron reservoirs.
In slave boson representation \cite{barnes}, our Hamiltonian reads
\begin{eqnarray}
H&=&\sum _{p,\sigma,\alpha}(\epsilon _p+\mu _{\alpha} +\Delta^{\alpha}(t))
c^{\alpha \ \dagger}_{p\sigma}  c^{\alpha}_{p\sigma}+
(\epsilon _d + \Delta (t)) \sum _{\sigma} f^{\dagger }_{\sigma } f_{\sigma}
\nonumber \\
&+& \sum _{p,\sigma,\alpha} T_{\alpha}
(f^{\dagger }_{\sigma} b
c^{\alpha}_{p\sigma} + h.c.) \label{1}
\end{eqnarray}
where the first term describes the noninteracting electrons
of the left and right lead with different time dependent
chemical potentials $\mu _{\alpha} + \Delta^{\alpha}(t)$,
$\alpha=L,R$ (we take $\mu_{\alpha}$ to be
+/- V/2, respectively, for the finite bias case).
The second and third terms describe
the time dependent dot level $\epsilon _d + \Delta (t)$ and the
tunneling terms, respectively, where f and b are the slave fermion
and slave boson operators. The physical particle operator on
the dot is represented by $d^{\dagger}_{\sigma}=
f^{\dagger}_{\sigma} b$, supplemented by the constraint
$\sum _{\sigma } f^{\dagger}_{\sigma}f_{\sigma} + b^{\dagger}b =1$.
With a standard canonical transformation we can move all explicit time
dependence
to the tunneling matrix elements $T_{\alpha}\rightarrow T_{\alpha}(t)$.
We have
$T_{\alpha} (t) = T_{\alpha} exp \left( i \int_{-\infty}^t d\tau [
\Delta^{\alpha}(\tau) - \Delta(\tau)] \right)$,
where $\Delta^{\alpha}(\tau), \Delta(\tau)$ are the time variations
of the energy levels in the leads and the dot, respectively
(throughout this work we set $\hbar=1$. From
this we can see that only the time dependent motion of
the leads relative to the dot is important.
For numerical convenience, we assume that $\Delta^{\alpha}(\tau) -
\Delta(\tau)$ takes a simple form, namely
$\Delta^{\alpha}(\tau) - \Delta(\tau) = \Delta_\alpha
\mbox{sin}\Omega\tau$, that is, a harmonic oscillation with (outer)
frequency $\Omega$ and amplitude $\Delta_{\alpha}$.

We compute the differential conductance within the
Non Crossing Approximation (NCA) for the Anderson model
in the Kondo limit\cite{muhart,cole,bickers}.
The NCA has been very successful in describing the equilibrium Kondo problem
except for the appearance of spurious nonanalytic behavior at a
temperature far below the Kondo temperature $T_K$.
These spurious low-T properties are due to the fact that the  NCA neglects
vertex
corrections responsible for restoring the low $T$ Fermi liquid
behavior \cite{costi}.
However, there is a wide range of temperature where $T$ is
much higher than the energy scale of these spurious effects and still well
below $T_K$. For quantum dots, we are interested in temperatures of about
$T_K$, since smaller temperatures are very hard to reach experimentally.
Therefore, the NCA should give a reliable  qualitative understanding
of the physics in typical quantum dots.

In order to calculate the conductance in nonequilibrium, the NCA must be
generalized using nonequilibrium Green functions
\cite{langreth}. One solves for both
the retarded Green functions for these operators,
$G^r$ (fermions) and $D^r$ (boson), and for the `lesser' Green functions
$G^<$ and $D^<$,
which contain information about the nonequilibrium distribution function.
The derivation of the integral equations for these four functions
follows the work of Meir, Wingreen, and Lee \cite{mei-win-lee}, but has to be
generalized to the case of time dependent perturbations
(for the adiabatic case see Ref.~\cite{sha-lan-nor}).
One obtains complicated equations
for the self-energies $\Sigma$ (fermion) and $\Pi$ (boson)
(lesser, retarded) which now (aside from the usual
frequency dependence) have an additional (discrete) index $n$ referring to the
harmonics due to the explicit time dependence of the hamiltonian.
The equations (which we can not show due to restricted space) are very
hard to solve in their full generality, since they involve three constrained
sums over the new index $n$.
However, in the nonadiabatic regime, where
the outer frequency $\Omega$ is much larger than the Kondo temperature
$T_K$ (or the tunneling rates $\Gamma_\alpha
(\omega)=2\pi T^2_\alpha N_\alpha(\omega)= \Gamma_\alpha N_\alpha(\omega)$
in the mixed valence regime
($N_\alpha(\omega)$ is the density of states in reservoir $\alpha$))
, we can restrict
ourselves to the dc \rm- components of the Green functions
and the self energies, since the ac--components are suppressed by
a factor $n T_K/\Omega$ ($n \Gamma_\alpha /\Omega$ in the mixed valence
regime). Then the equations reduce to
\begin{mathletters}
\label{2}
\begin{eqnarray}
\Sigma^<_o(\omega) = \frac{ 1}{2\pi} \int d\omega ' K^+(\omega -\omega ')
D^<_o(\omega ')\,\, , \\  \Pi^<_o(\omega) = \frac{ N}{2\pi} \int d\omega '
K^-(\omega ' -\omega ) G^<_o(\omega ')
\end{eqnarray}
\end{mathletters}
with the integration kernels given by $K^\pm=\sum_\alpha K^\pm_\alpha$ and
$K^{\pm}_\alpha(\omega)=\int d\omega ' \gamma^\pm_\alpha (\omega ')
P_\alpha (\omega - \omega ')$. Here $\gamma^+_\alpha
(\omega)=\Gamma_\alpha (\omega) f_\alpha (\omega)$ and
$\gamma^-_\alpha (\omega)=\Gamma_\alpha (\omega) (1-f_\alpha(\omega))$
are the transition rates in lowest order for an electron to enter
(leave) the quantum
dot in the absence of time-dependent perturbations ($f_\alpha
(\omega)=(\exp{\beta (\omega-\mu_\alpha)} +1)^{-1}$ is the Fermi
distribution function for the lead $\alpha$). Furthermore,
\begin{equation}\label{3}
P_\alpha (\omega)=\sum_n J^2_n ({\Delta_\alpha\over\Omega})\delta
(\omega - n\Omega)
\end{equation}
denotes the probability for an electron in reservoir
$\alpha$ to emit or absorb the energy $\omega$. Since we are
considering microwave irradiation this energy has to be identical
to multiples of $\Omega$. This reflects the physical effect that
electrons in the leads can now absorb or emit energy quanta
$n\Omega$ which gives rise to a shift of the whole Fermi sea.

In principle, the sum over $n$ in Eq.~(\ref{3}) runs from $-\infty$ to
$\infty$. However, the
higher Bessel functions have negligible values (compared to the $n=0,1,2,3$
Bessel functions) for arguments $\Delta_{\alpha}/\Omega < 3$, especially since
they enter the integration kernels squared. Therefore, if we restrict
ourselves to this range of argument, we can in good approximation cut
off the sum at $n=2$.

Furthermore we find the relation
\begin{equation} \label{4}
G^<_o(\omega)  = \Sigma_o^<(\omega)\, |G^r_o(\omega)|^2 \,\, ,
\,\,\,
D^<_o(\omega)  = \Pi_o^<(\omega)\, |D^r_o(\omega)|^2 \,\, .
\end{equation}
Thus, we need the retarded Green functions as input in order to solve the
equations for the lesser Green functions. They can be found from
$G^r_0(\omega)=(\omega-\epsilon_d-\Sigma^r_0(\omega))^{-1}$,
$D^r_0(\omega)=(\omega-\Pi^r_0(\omega))^{-1}$ and
\begin{mathletters}
\label{5}
\begin{eqnarray}
Im\Sigma^r_o(\omega) = \frac{1}{2\pi} \int d\omega ' K^-(\omega -\omega ')
Im D^r_o(\omega ') \,\, ,\,\,\,\, \\
Im\Pi^r_o(\omega) = \frac{N}{2\pi} \int d\omega ' K^+(\omega ' -\omega )
Im G^r_o(\omega ') \,\,\,\,\,\,.
\end{eqnarray}
\end{mathletters}
The real parts of the retarded self energies can be found from the
Kramers--Kronig relation.

We have solved Eqs.~(\ref{2}), (\ref{4}) and (\ref{5}) numerically by
iteration. The code is designed to
sacrifice memory economy for higher speed. We are able to go to temperatures
orders of magnitude below $T_K$, although for the cases of interest here
it is enough to go to $T_K$/5 (where the Kondo peaks have essentially reached
their low temperature form). The validity of the numerical solution is
established by checking the sum rules. They are fulfilled to within 0.5\%
even in
the worst cases (lowest T, large $n$, etc.). The constraint is enforced
after each iteration, so that no unphysical solution can emerge.

Aside from the density of states of the lead electrons, all the integration
kernels can be interpreted as modified distribution functions with steps
heights (at low temperatures) determined mostly by the Bessel functions.
Since the dot spectral function tends to have a peak at energies close
to these
steps (if they are pronounced enough), we anticipate a splitting of the
equilibrium Kondo peak into a variety of smaller peaks at positions determined
by  $ \pm n\Omega \pm V/2$.

The  dc \rm- dot spectral function, $A_d$, is computed from the
slave Green functions via the convolution \cite{het-kro-her}
(using the conventions of
M\" uller--Hartmann \cite{muhart} for the spectral functions,
$A (\omega )= -\mbox{Im}G_o^r(\omega )/\pi$,
$B(\omega )= -\mbox{Im}D_o^r(\omega )/\pi$,
and the `lesser' Green functions,
$a(\omega )= \mbox{Im}G_{o}^{<}(\omega )/2\pi$ and
$b(\omega )= \mbox{Im}G_{o}^{<}(\omega )/2\pi$),

\begin{eqnarray}\label{6}
 A_{d}(\omega) = \int \frac{d\epsilon}{\pi}
\left[\, a (\epsilon) B (\epsilon -\omega) + A (\epsilon) b (\epsilon -\omega)
\right] .
\end{eqnarray}
Generalizing the known formulas \cite{her-dav-wil,mewin2} we can
calculate the dc \rm current from
$A_{d}(\omega)$ (if the couplings to the left and right leads are the same),
\begin{eqnarray}\label{7}
I=  \frac{eN}{2} \int d\omega A_{d}(\omega)
\left[K^+_L(\omega)-K^-_R(\omega)\right]
\label{curr}
\end{eqnarray}
and from the current  the conductance $G(V) = d I(V)/d V$ via numerical
derivative.
We can also calculate the magnetic susceptibility and other transport
properties. Additionally, it is possible to apply this method to bulk Kondo
systems
in the presence of time dependent fields. Results on that and a more detailed
discussion of the full NCA equations, approximation and the numerics will be
published elsewhere.

Since the Anderson model in the presence  of a static potential (finite bias)
alone has been discussed in \cite{mei-win-lee} and for the two channel
case in  \cite{het-kro-her} we will start the discussion of results
for the zero bias case with an oscillating  potential with frequency
$\Omega$ and amplitude $\Delta_L =\Delta_R=\Delta$ ($\Delta/\Omega =
1.4$).

Taking the Kondo limit with $\epsilon_d$ large and negative
($\epsilon_d = - 1340 T_K$, except for Fig. \ref{covsed}) we obtain a
resonance (Kondo
peak) with a width at  half maximum (up to a factor of the order
of unity) given by $T_K$ \cite{bickers}.
Applying now an oscillating potential of frequency $\Omega =  16 T_K$
the Kondo peak splits, creating side peaks at +/- $n\Omega$
shifted frequencies, see the top curve of Fig. \ref{diva}.
For $\Omega$ much below that value our approximation might not be valid.
If we nevertheless look at the behavior of the Kondo peak for these
smaller values
we find that the original peak first broadens and finally splits into three
peaks
(in analogy to the case of a static potential where it splits into two peaks
\cite{mei-win-lee,het-kro-her}). Observe the asymmetry
of the side peaks, the one at positive frequencies is higher than the one at
negative frequencies as long as the negative one is not merging with the
broad hump (with width $N \Gamma = N (\Gamma_L +\Gamma_R), \Gamma = 300 T_K$
except for Fig. \ref{covsed})
of the bare dot level (not shown, since far below the Fermi level).

This asymmetry can be used to built a 'Kondo pump'. Suppose the right lead
is kept at a fixed potential (Fermi energy) whereas the left one has the
oscillation with frequency $\Omega$ applied, that is $\Delta_L = \mbox{finite},
\,\Delta_R = 0$ . In the left lead electrons in an energy range up to $\Omega$
below the Fermi energy can be excited and move to the right lead via
the states at the dot with energies above the Fermi level. On the other
hand electrons in the right can move to the left lead using states of
the dot with energies below the Fermi level. Due to the asymmetry of
the dot spectral function (which is enhanced upon what is seen in Fig.
\ref{diva} with symmetric coupling) these currents will not be equal.
Therefore,
one should observe a \bf dc \rm current through the dot, even if no finite
bias $V$
is applied. This current will be directed from left (right) to right (left)
if $\Delta_L > (<) \Delta_R$. In Fig. \ref{asymm}
we show a graph of this 'pump'
current in dependence of $\Delta_R$ for a fixed value of $\Delta_L$.
The current is quite large  (equivalent to a bias $V$ of about 1/10 $T_K$
if just a static potential is applied with symmetric couplings)
due to peaks in the
dot spectral function at the 'right' frequencies. At these frequencies the
lorentzian spectral function of the corresponding
noninteracting resonant level model would be unspectacular flat and very small
for the given set of parameters of dot level energy $\epsilon_d$ and width
$\Gamma$. This current
enhancement is strong temperature dependent and one might naively expect
that it vanishes as $T$ becomes larger than $T_K$.
However, as long as $T$ is well below $\Omega$  the effect remains,
since the equilibrium $T_K$ is no longer the relevant width of the Kondo
peaks \cite{mei-win-lee,het-kro-her}.

Another effect partly due to the asymmetry of $A_d$ is an asymmetry of the
zero bias conductance (now again $\Delta_L = \Delta_R = \Delta$, with
$\Delta/\Omega
= 1.4$, so that only the zeroth and first Bessel function have
relevant effects on $A_d$) as function of the gate voltage
(Coulomb-oscillations). We increase the level energy $\epsilon_d$
from the Kondo limit through the mixed valence regime into the empty dot
regime (where $\epsilon_d$ is now large and positive). Taking $\Omega > \Gamma$
($T << \Gamma$), we obtain a peak in the conductance for $\epsilon_d$
slightly below
$- \Omega$, the usual, higher peak just below the Fermi energy,  and another,
even higher peak slightly below $+ \Omega$, all with the bare level width
$\Gamma$ as shown in Fig. \ref{covsed}. This asymmetry in the heights of the
conductivity peaks can be easily understood by considering  the increase in
total spectral weight $\int d\omega A_d (\omega)$. Having a value of just
below 1/2 (per spin) in the Kondo limit ($\epsilon_d$ large and negative),
it increases monotonically and approaches unity in the empty dot limit.
The increase in total weight is rising the value of $A_d$ at all frequencies,
leading to an increased conductivity peak height for the $+\Omega$ shifted
peak. This is again in contrast to the resonant level case, where the total
weight of the lorentzian spectral function remains constant (equal unity)
which leads therefore to equal heights for corresponding side peaks.
%

We  can also study the influence of the outer frequency $\Omega$ on the
temperature dependence of the zero bias conductance. In equilibrium,
the zero bias conductance saturates for $T < T_K$ and drops logarithmically
for $T >> T_K$, see Fig. \ref{condy}.
However, for $\Omega = 10 T_K$ the main drop now occurs
somewhat below $\Omega$, with $T_K$ playing no significant role anymore.
The low $T$ zero bias conductance depends roughly logarithmically on $\Omega$
in the nonadiabatic regime $\Omega > 10 T_K$. At the lowest $\Omega$ for which
we can trust our approximation the zero bias conductance is already
suppressed by about one order of magnitude.

\begin{figure}
\caption{Nonlinear conductance $G(V)$ vs. bias $V$ for an outer
frequency $\Omega
= 16 T_K$. $G(0)$ denotes the zero bias conductance $G(V=0,T)$.
Aside from the usual zero bias peak (not fully resolved) there
are peaks just below multiples of $\Omega$. The first peak is lower
relative to the background than the second peak. The peaks are due to
merging of side peaks at the corresponding bias $V$ (see Fig. 4).
These conductance peaks are present even though $T$ is larger $T_K$ as long as
$T << \Omega << \Gamma$.}
\label{conduc}
\end{figure}
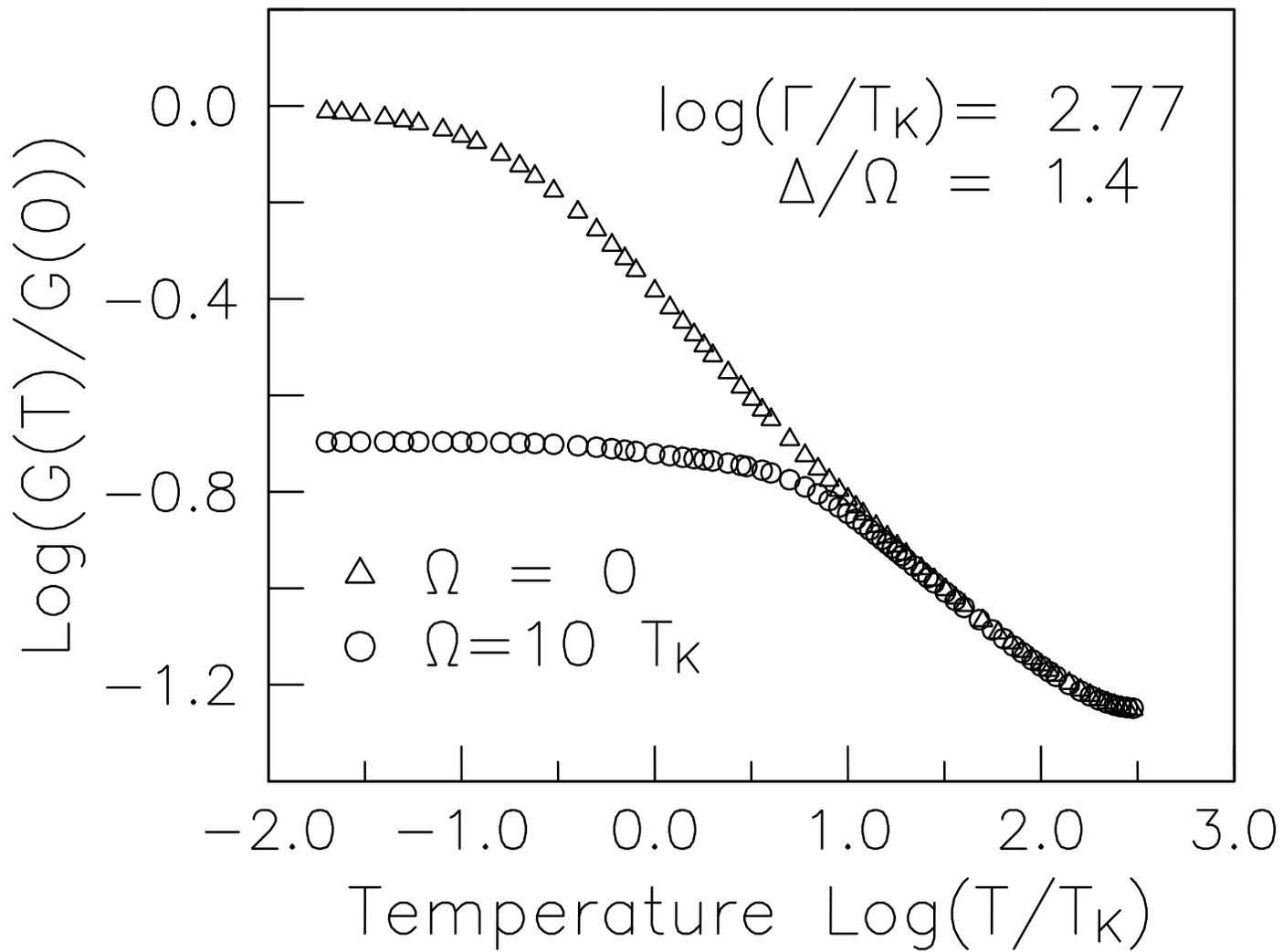
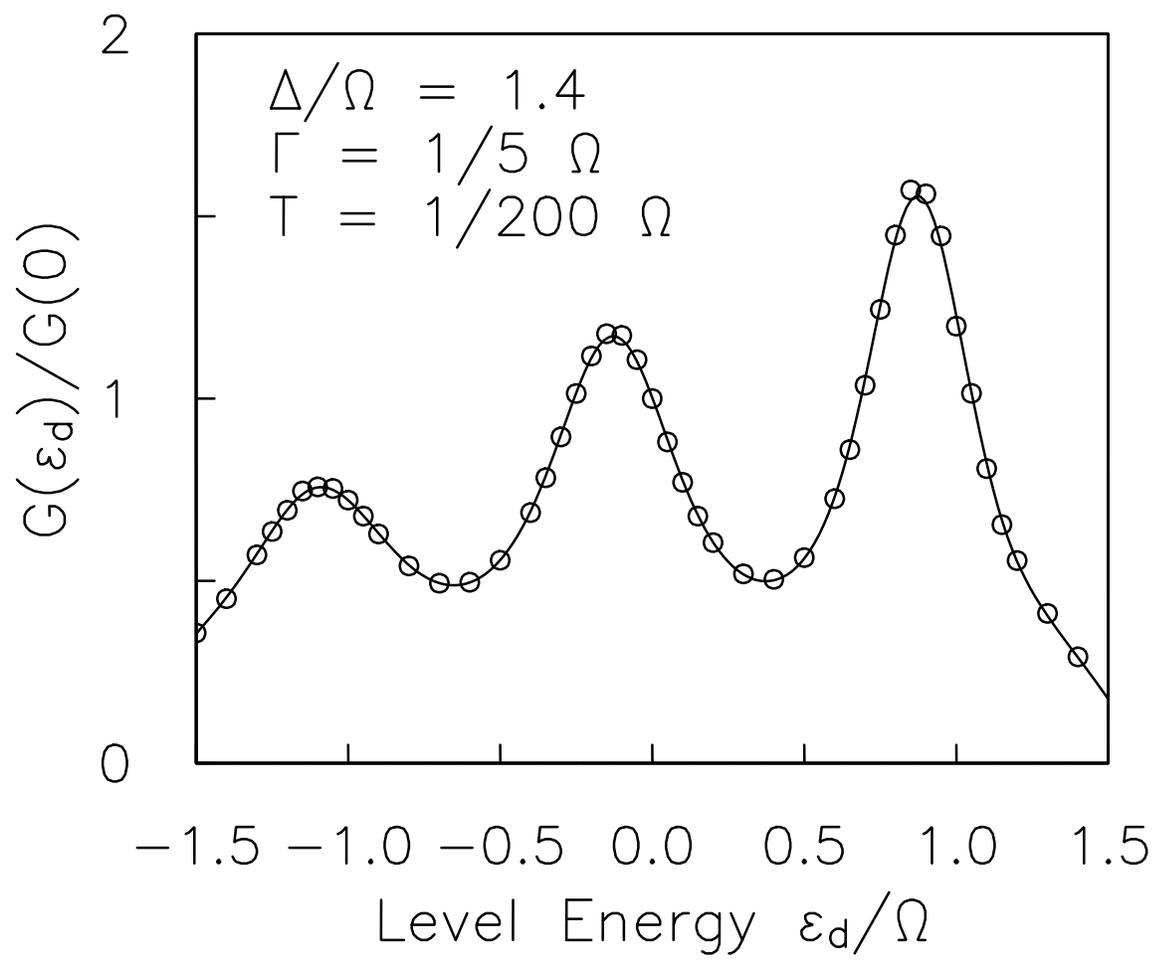
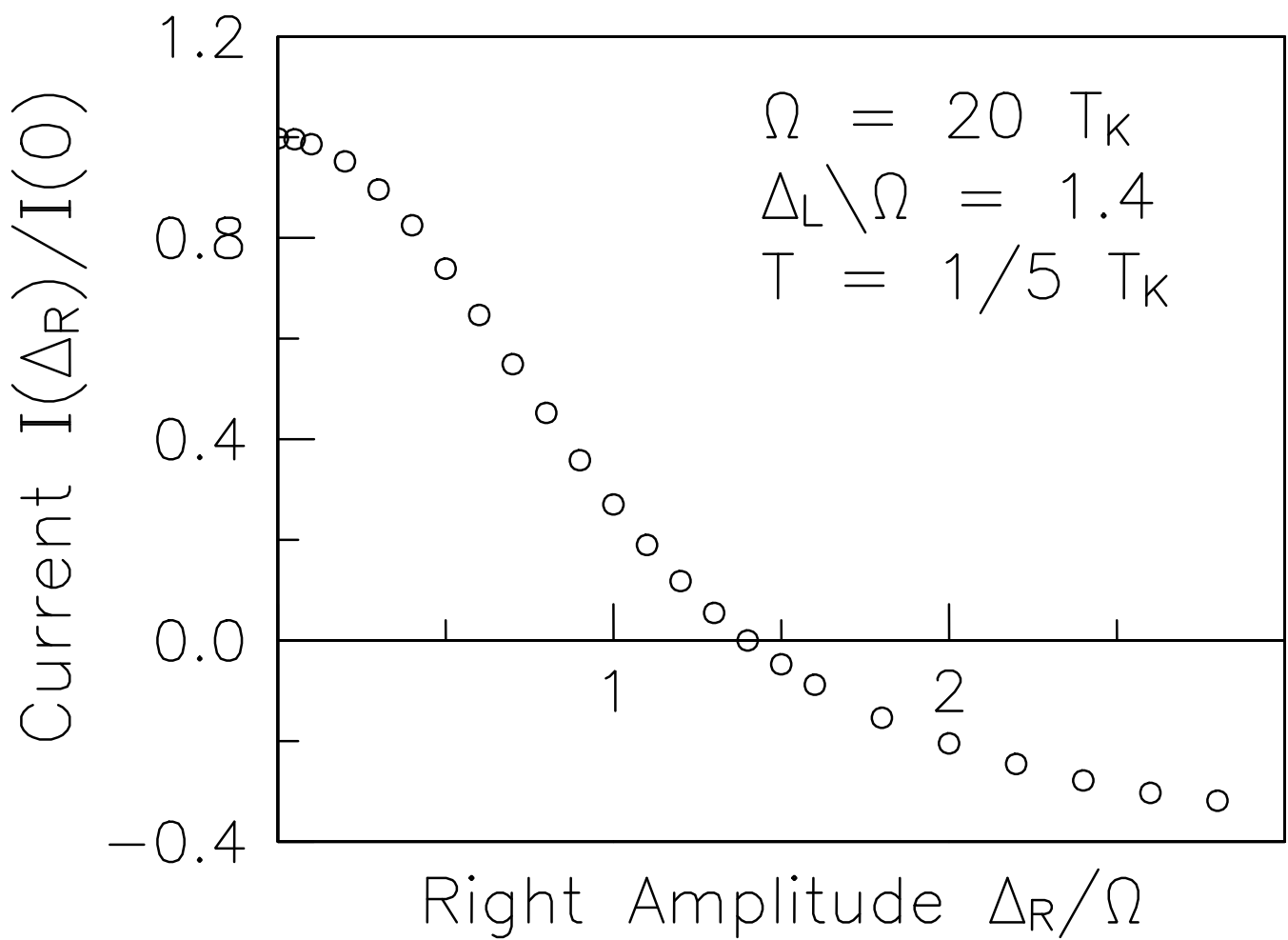
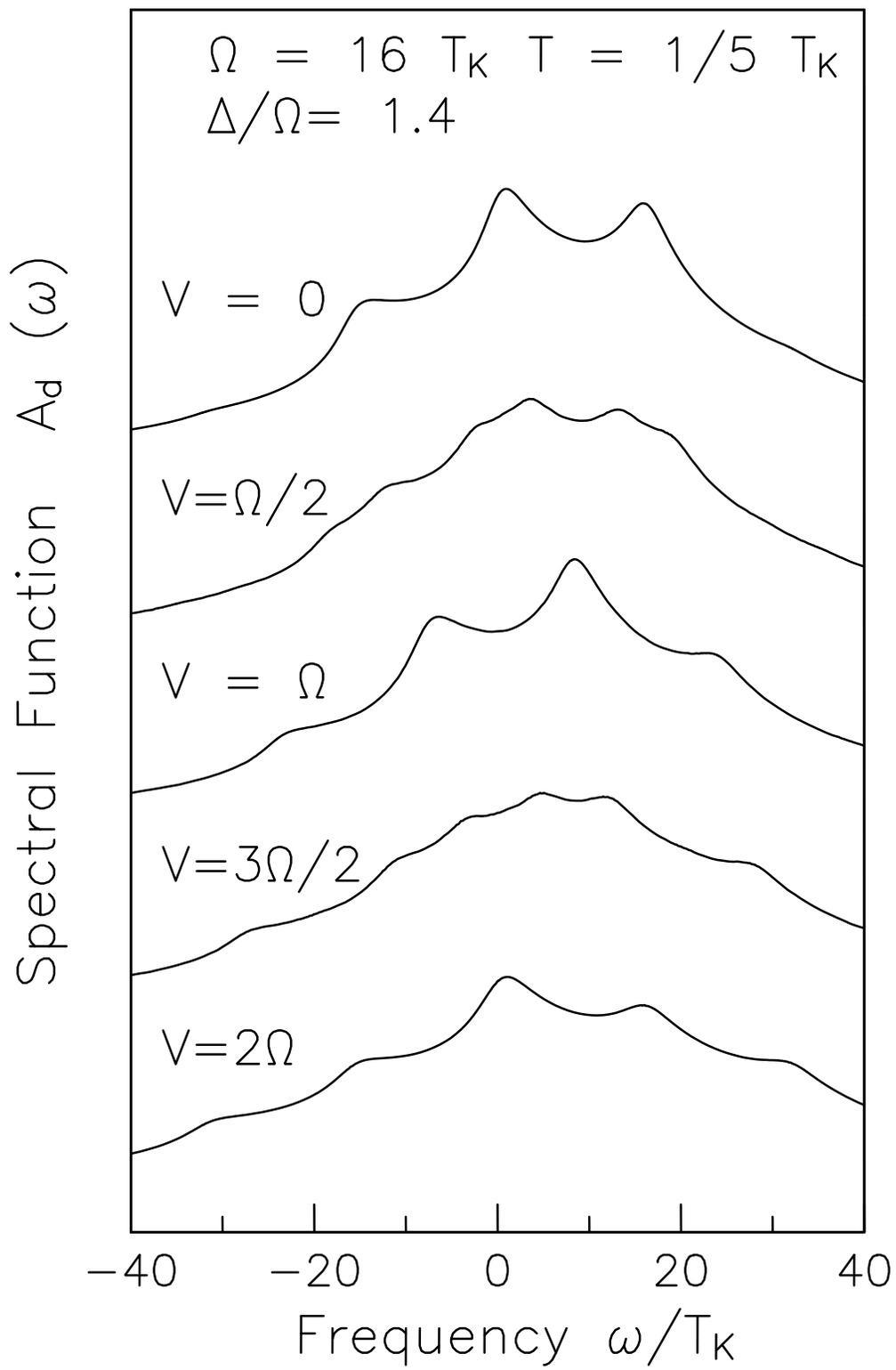
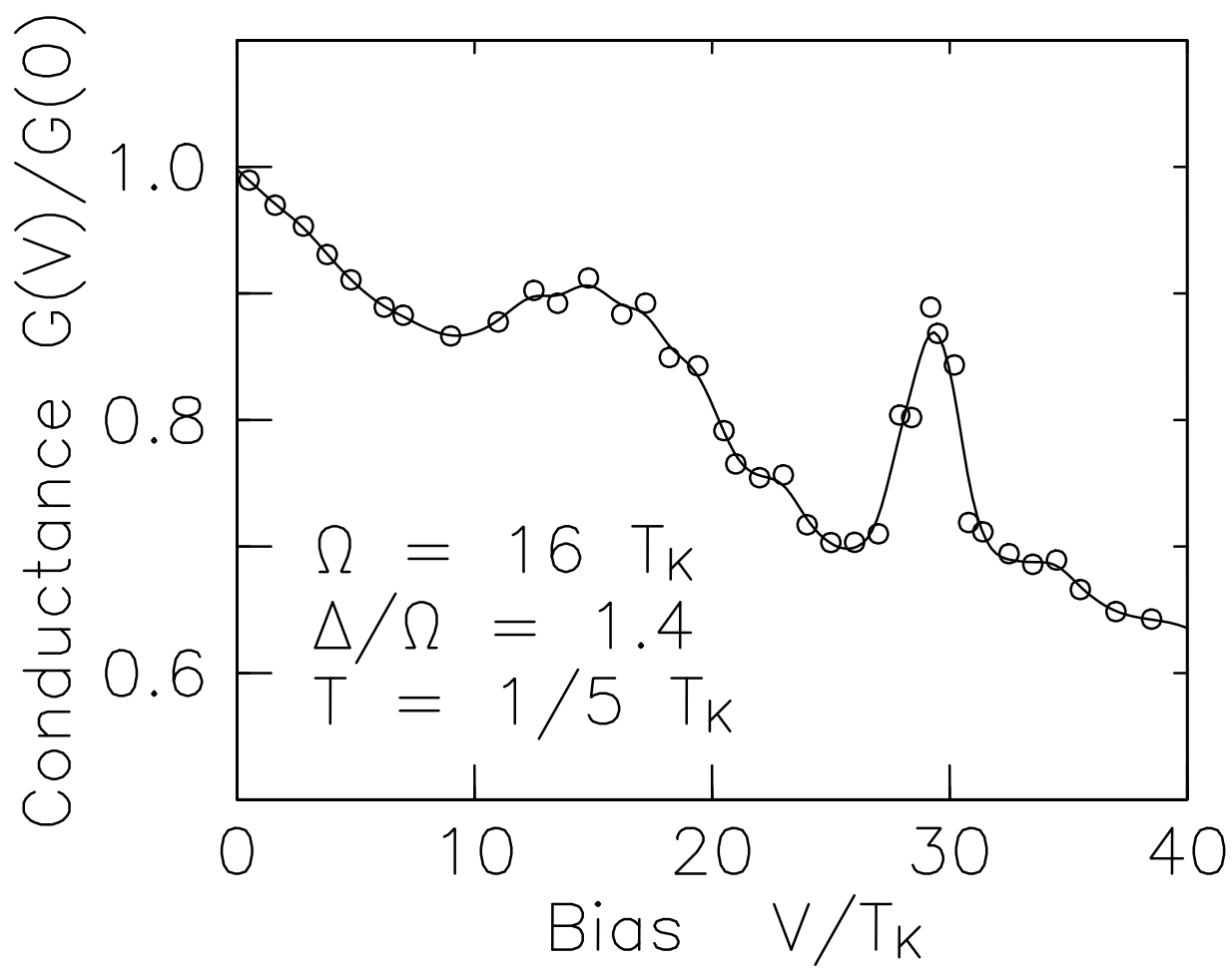

In addition, we now apply a finite bias $V$ to the dot. If $V$ and $\Omega$
are large enough (and $T$ low enough)
to allow a clear separation of all features, the applied
bias splits all peaks into two (see Fig. \ref{diva}), as it did in the static
case ($\Delta = 0$)
with the single Kondo peak \cite{mei-win-lee,het-kro-her}. If $V$ is now
increased until
it is close to $\Omega$, the peaks positioned at $-\Omega + V/2$ and
$-V/2$ as well as the peaks at $+V/2$ and $+\Omega - V/2$ merge. By further
increasing $V$ until it is close to $2 \Omega$, we reach the point where
the peaks positioned at $-\Omega +V/2$ and $+\Omega - V/2$ merge into one peak
at about the Fermi energy. It seems obvious that the current should increase
rapidly
around these special values of $V$, leading to strongly enhanced conductances.
However, due to broadening while increasing $V$, the overall values of
$A_d$ monotonically decrease, and spectral weight is just shifted from the
valleys to the peaks when these mergings of peaks occur. The final say in this
matter has the numerical evaluation of the conductance. Fig. \ref{conduc}
shows the
conductance $G(V)$ for a low temperature $T = T_K /5$. Indeed there are
strong features in the conductance at values of $V$ slightly below the
naively expected values. Especially the peak at $2 \Omega$ is quite
distinguished (it rises 25--30 \% above its background) and the width is
in rough agreement with the width of the peaks of $A_d$ at this bias.
The first feature
is less pronounced and wider, which we can trace back to uncomplete separation
before the merging of the peaks at this lower value of $V$. Again, this
feature is strongly $T$-- dependent, but, as long as $T$ is low enough to
allow the separation and merging of peaks without washing them out altogether
(roughly $T \sim 1/10 \Omega$),
similar, though wider and smaller conductance anomalies should 	be observable.
%
%

In conclusion, we performed numerical calculations which constitute the first
theoretical work on the Anderson model with infinite strong Coulomb repulsion
at low temperatures in the presence of time dependent
perturbations. We compute the conductance in the linear and nonlinear
response regime. Both show peaks in accordance to
theoretical arguments. We find the effect of a 'Kondo pump' which
creates a dc \rm current in the presence of an ac \rm
potential alone, if the
amplitudes of potential oscillations in the leads are different from each
other. All these features may be used to experimentally clarify the question of
the presence of Kondo physics in quantum dots. The temperature dependence of
these effects are strong, but observability is not limited to temperatures
below the equilibrium Kondo temperature $T_K$.

Acknowledgements: We would like to thank S. Hershfield, J. Kroha, P.J.
Hirschfeld, A. Schiller, H.v. L\"ohneysen and P. W\"olfle
for useful discussions.
This work was supported by NSF grant DMR9357474, the U.F.~D.S.R.,
and the NHMFL (M.H.), as well as the Swiss National Science Foundation
and the ''Deutsche Forschungsgemeinschaft'' as part of
''Sonderforschungsbereich 195'' (H.S.).

\begin{figure}
\caption{Spectral function $A_d$ for $\Omega = 16 T_K$ vs.
frequency for various
applied bias $V$ (curves are offset). The topmost curve shows the splitting
of the Kondo peak into three peaks at $-\Omega, 0 (= \epsilon_F)$ and
$+\Omega$ at zero bias.
The other curves illustrate the merging of peaks at the
voltages $V = \Omega$ and $V = 2\Omega$. Also observe the general broadening
of the peaks upon increase of bias $V$.}
\label{diva}
\end{figure}
\begin{figure}
\caption{'Pump'--Current $I(\Delta_R)$ vs. the amplitude of oscillations
$\Delta_R$ in
the right lead.
For fixed left amplitude $\Delta_L$ the current is nonzero for all values
of $\Delta_R \neq \Delta_L$ even though no bias $V$ is applied. $I(0)$ denotes
$I(\Delta_R = 0)$.
This 'Kondo pump' effect can be explained with the asymmetry
of the split Kondo peaks about the Fermi level.}
\label{asymm}
\end{figure}
\begin{figure}
\caption{Zero bias conductance $G(\epsilon_d)$ vs. dot--level energy
$\epsilon_d$. The three
peaks slightly below $-\Omega, 0$ and $+\Omega$ show an increasing height
mostly due to the increasing spectral weight when $\epsilon_d$ moves
from the Kondo limit to the empty dot limit. $G(0)$ denotes
$G(\epsilon_d = 0)$.}
\label{covsed}
\end{figure}
\begin{figure}
\caption{
Zero bias conductance $G(T)$ vs. temperature $T$ on double--logarithmic scale.
Shown are data for the equilibrium $\Omega = 0$ and for $\Omega =
10 T_K$. $G(0)$ denotes $G(T=0)$ for the case ($\Omega = 0$).
In equilibrium the zero bias conductance starts to drop already below $T_K$.
With an outer frequency $\Omega$ applied the main drop
starts just below the corresponding temperature, the equilibrium $T_K$ playing
only a minor role. At very low temperatures $G(T)$ is
reduced by almost one order of magnitude for this value of $\Omega$.
At high T all features due to $\Omega$
are washed out and $G(T)$ is independent of $\Omega$.}
\label{condy}
\end{figure}
\end{document}